\documentclass[]{aa}

\usepackage{graphicx}
\usepackage{color}

\begin{document}

\title{A new interpretation of the far-infrared - radio correlation and the expected breakdown at high redshift}

\titlerunning{The far-infrared - radio correlation at high redshift}
\authorrunning{Schleicher \& Beck}

\author
  {Dominik R.G. Schleicher
  \inst{1}
  \and
  Rainer Beck
  \inst{2}
  }

\institute{Institut f\"ur Astrophysik, Georg-August-Universit\"at G\"ottingen, Friedrich-Hund-Platz 1, 37077 G\"ottingen, Germany \\
\email{dschleic@astro.physik.uni-goettingen.de}
\and
Max-Planck-Institut f\"ur Radioastronomie, Auf dem H\"ugel 69, 53121 Bonn, Germany\\
\email{rbeck@mpifr-bonn.mpg.de}
}

\date{\today}


\abstract
{Observations of galaxies up to $z\sim2$ show a tight correlation between far-infrared and radio continuum emission, suggesting a relation between star formation activity and magnetic fields in the presence of cosmic rays.}
{We explain the far-infrared - radio continuum correlation by relating star formation and magnetic field strength in terms of turbulent magnetic field amplification, where turbulence is injected by supernova explosions from massive stars. We assess the potential evolution of this relation at high redshift, and explore the impact on the far-infrared -- radio correlation.}
{We calculate the expected amount of turbulence in galaxies based on their star formation rates, and infer the expected magnetic field strength due to turbulent dynamo amplification. We calculate the timescales for cosmic ray energy losses via synchrotron emission, inverse Compton scattering, ionization and bremsstrahlung emission, probing up to which redshift strong synchrotron emission can be maintained.}
{We find that the correlation between star formation rate and magnetic field strength in the local Universe can be understood as a result of turbulent magnetic field amplification. The ratio of radio to far-infrared surface brightness is expected to increase with total field strength. A continuation of the correlation is expected towards high redshifts. If the typical gas density in the interstellar medium increases at high $z$, we expect an increase of the magnetic field strength and the radio emission, as indicated by current observations. Such an increase would imply a modification, but not a breakdown of the far-infrared -- radio correlation. We expect a breakdown at the redshift when inverse Compton losses start dominating over synchrotron emission. For a given star formation surface density, we calculate the redshift where the far-infrared -- radio correlation will break down, yielding $z\sim(\Sigma_{\mathrm{SFR}}/0.0045\ M_\odot\ \mathrm{kpc}^{-2}\ \mathrm{yr}^{-1})^{1/(6-\alpha/2)}$. In this relation, the parameter $\alpha$ describes the evolution of the characteristic ISM density in galaxies as $(1+z)^\alpha$. We note that observed frequencies of $1-10$~GHz are particularly well-suited to explore this relation, as bremsstrahlung losses could potentially dominate at low frequencies.}
{Both the possible raise of the radio emission at high redshift and the final breakdown of the far-infrared -- radio correlation at a critical redshift will be probed by the Square Kilometre Array (SKA) and its pathfinders, while the typical ISM density in galaxies will be probed with ALMA. The combined measurements will thus allow a verification of the model proposed here.}

\maketitle


\section{Introduction}
While non-thermal radio emission arises due to synchrotron emission of cosmic ray electrons in the interstellar magnetic fields, far-infrared radiation is 
generated by dust grains heated by UV radiation from massive stars, which typically dominate the Lyman continuum luminosity. A correlation between the observed fluxes from these two phenomena has been observed by \citet{deJong85} and \citet{Helou85}, which became known as the far-infrared -- radio correlation. As the radio emission requires the presence of cosmic rays and magnetic fields, while far-infrared emission requires active or very recent star formation, the presence of this relation has been interpreted by \citet{NiklasBeck97} in terms of a relation between magnetic field strength, gas density and star formation rate. While their observations showed a relation between magnetic field strength and gas surface density, it is well-known that the gas surface density also correlates with star formation activity, known as the Kennicutt-Schmidt-relation \citep{Schmidt59, Kennicutt98, Kennicutt08, Walter08, Kennicutt12}.

The far-infrared -- radio correlation has been tested in many studies both concerning global galaxy properties in the local Universe \citep[e.g.]{Niklas97, Yun01}, as well as on local scales employing high-resolution observations of individual galaxies \citep{Beck88, Marsh95, Hoernes98, Chyzy07, Dumas11, Basu12}. It is particularly remarkable that the correlation holds even for small galaxies like the Magellanic Clouds (LMC, SMC) \citep{Xu92, Hughes06, Leverenz13}. A more recent study shows the role of the magnetic fields in controlling the radio-infrared correlation in nearby galaxies \citep{Tabatabaei13}.

In order to advance its physical interpretation, \citet{Murphy06, Murphy08} employed a cosmic-ray diffusion model, which allowed them to derive the expected radio maps from existing far-infrared observations, yielding a good agreement with the observed data. The transport of cosmic rays in the interstellar medium is indeed frequently modeled as a diffusion process \citep{Ginzburg80}, and diffusion coefficients have been obtained both for the solar neighborhood \citep{Jones01, Moskalenko02, Maurin02} as well as for extragalactic systems \citep{Dahlem95, Heesen09, Heesen11, Murphy12, Buffie13, Tabatabaei13b}.

The origin of this relation is likely related to the prevalent energy-loss mechanisms for cosmic-ray electrons, such as synchrotron radiation and inverse Compton scattering \citep{Condon92, Thompson06, Lacki10}, escape of cosmic rays from the galaxy \citep{Helou93} and potentially additional processes like ionization or bremsstrahlung \citep{Longair94, Murphy09}. Due to the underlying complexity of the physical situation, it is thus surprising that a well-defined correlation emerges over nearly five orders of magnitude in luminosity \citep[e.g.][]{Yun01}. As noted above, this relation likely reflects an underlying relation between star formation and magnetic fields, which has been recently confirmed even for dwarf irregulars by \citet{Chyzy11} and \citet{Roychowdhury12}.

In order to understand this relation in the context of galaxy evolution, it is  central to test this relation even at high redshift. A first study by \citet{Garrett02} was based on a sample in the HDF-N with a redshift median of $0.7$, finding that the far-infrared -- radio correlation is valid in this regime. \citet{Gruppioni03} independently confirmed the relation out to $z\sim0.6$, while \citet{Appleton04} report strong consistent evidence for this relation out to $z\sim1$. Based on the Herschel-ATLAS survey, \citet{Jarvis10} reported an in-depth study of the relation out to $z\sim0.5$, while \citet{Sargent09} and \citet{Bourne11} explored it out to $z\sim2$ in the COSMOS field and for massive galaxies, respectively. So far, the high-redshift observations  indicate no strong evolution with redshift. We note that, in a compilation by \citet{Murphy09}, some $z\sim2.2$ and $z\sim4.4$ submillimeter galaxies lie below the canonical ratio by about a factor of $2.5$, implying either reduced far-infrared or enhanced radio emission for these galaxies. More recently, \citet{Ivison10a} and \citet{Ivison10c} probed the evolution of \begin{equation}
q_{\mathrm{IR}}=\log_{10}\left[ \left( \frac{S_{\mathrm{IR}}}{3.75\cdot10^{12} W m^{-2}}  \right) / \left( \frac{S_{1.4\ \mathrm{GHz}}}{Wm^{-2}\mathrm{Hz}^{-1}}  \right) \right]
\end{equation}
with Herschel, finding that it evolves as $q_{\mathrm{IR}}\propto(1+z)^\gamma$, with $\gamma\sim0.26\pm0.07$. The latter indicates an increasing radio or a decreasing infrared emission with higher redshift. { Similar results have been reported by \citet{Casey12}, employing Keck spectroscopic observations of 767 Herschel-SPIRE selected galaxies, emplying an evolution as $q_{\mathrm{IR}}\propto(1+z)^{-0.30\pm0.02}$ at $z<2$. The evolution of the infrared luminosity function at high redshift is probed in further detail by \citet{Magnelli11} and \citet{Barger12}. In spite of potential uncertainties such as AGN contamination \citep{Daddi07, DelMoro13}, the results suggest an enhanced radio flux or a decreasing far-infrared emission at high redshift.} While an increase of the radio emission may not be expected at first sight, we will discuss in this paper how it can be linked to the evolution of the typical densities in the interstellar medium (ISM).

From a theoretical point of view, these results are in fact not surprising. Efficient magnetic field amplification in the presence of turbulence, the so-called small-scale dynamo, has already been established by \citet{Kazantsev68} in the presence of subsonic Kolmogorov turbulence. Our theoretical understanding of this process has been extended in many subsequent studies, for instance by considering the effect of helicity \citep{Subramanian98, Subramanian99, Boldyrev05, Brandenburg05, Malyshkin07, Malyshkin09} and different types of turbulence \citep{Schober12b, Schober12c, Bovino13}.

 A particular concern are the effects of high Mach numbers, as expected in situations where turbulence is driven by supernova explosions. The first simulations exploring Mach numbers above one have been pursued by \citet{Haugen04c}, although the first systematic study exploring Mach numbers from $\sim0.01$ to $20$ was only recently pursued by \citet{FederrathPRL}, deriving the resulting growth rates and saturation levels. Most relevantly, their study demonstrated that the dynamo still exists at high Mach numbers, even though the saturation level is somewhat reduced. A similar behavior has been suggested based on analytical calculations by \citet{Schober12b}, which was extended into the non-linear regime where backreactions occur by \citet{Schleicher13}. A series of simulations by \citet{Balsara01, Balsara04, Balsara05} further demonstrated that the dynamo is  efficient in astrophysical environments when driven by supernovae.

Due to the rapid magnetic field amplification via turbulence, the small-scale dynamo was thus suggested to provide strong magnetic fields already during the formation of the first stars and galaxies \citep{Beck96, Arshakian09, Schleicher10c, deSouza10, Sur10, BeckA12, Sur12, Turk12, Schober12, Latif12}. More recently, turbulent magnetic field amplification was suggested to be efficient even in voids \citep{BeckA13}. While these systems are observationally out of reach, we can expect efficient magnetic field amplification in mature galaxies in the presence of turbulence. While these fields are initially not ordered, and dominate on scales comparable to the length scales of turbulence, they can nevertheless be detected via their resulting synchrotron emission. Indeed, as we propose in this paper, the magnetic field strength established via the small-scale dynamo may contribute strongly to the observed correlation between star formation rate and magnetic field strength in galaxies.

With the upcoming Square Kilometre Array (SKA)\footnote{SKA webpage: http://www.skatelescope.org/} and its Key Science Projects ``The Origin and Evolution of Cosmic Magnetism'' and ``Galaxy Evolution and Cosmology'', we can expect to probe magnetic fields \citep{Beck11} as well as the far-infrared -- radio correlation at higher and higher redshift \citep{Murphy09}, not only for the ultraluminous starbursts, but also for galaxies comparable to our Milky Way. Its precursors and pathfinders, including LOFAR\footnote{LOFAR webpage: http://www.lofar.org/}, MEERKAT\footnote{MEERKAT webpage: http://www.ska.ac.za/meerkat/} and ASKAP\footnote{ASKAP webpage: http://www.atnf.csiro.au/projects/mira/} are in fact expected to extend current measurements within a few years.

In this paper, we will formulate a theoretical model explaining the observed correlation between star formation rate and magnetic field strength in terms of turbulent magnetic field amplification, with turbulence injected by supernova explosions of massive stars. Based on this model, we can assess the potential evolution of this relation at high redshift. In order to explore whether also the far-infrared -- radio correlation will hold on an approximate level, we calculate the timescales of the characteristic energy losses of the cosmic ray electrons, assuming that the correlation is only established when synchrotron emission yields the shortest timescale. Based on this assumption, we derive a critical magnetic field strength which is required for efficient synchrotron emission at a given redshift,  corresponding to a critical star formation rate in our model. As a result, we obtain theoretical predictions up to which redshift the far-infrared -- radio correlation is expected to hold for a given star formation rate.

\section{An interpretation for the observed relation between star formation rate and magnetic field strength in galaxies}\label{interpret}
In this section, we propose an interpretation of the observed relation between star formation rate and magnetic field strength in galaxies in terms of turbulent magnetic field amplification, assuming that magnetic fields are predominantly amplified by supernova-driven turbulence. We derive here the expected correlation between star formation rate and magnetic field strength in such a model.  As shown by \citet{Thompson05}, the typical ISM density in galaxies can be parametrized as\begin{equation}
n=170\,\sigma_{200}^2 \, r_{\mathrm{kpc}}^{-2} \, Q^{-1}\ \mathrm{cm}^{-3},
\end{equation}
with $\sigma_{200}$ the stellar velocity dispersion in units of $200$~km/s, $r_{kpc}$ the scale under consideration and $Q$ the Toomre parameter describing gravitational (in)stability. They further show that the ratio of disk scale height to scale radius is given as\begin{equation}
\frac{h}{r}=f_g \, Q \, / \, 2^{3/2},
\end{equation}
with $f_g$ the gas fraction of the galaxy. While these quantities generally depend on the properties of the individual galaxy, we will in the following focus on  ``typical'' galaxies at a given redshift, implying a typical ISM density and a typical ratio $h/r$ at each redshift. The variations in these parameters are therefore expected to yield some scatter around the average relations derived here. We will however show below that the impact of such variations on our results is only small.

In the following, we therefore assume that the typical density in a galaxy depends only on redshift, and is given as\begin{equation}
\rho=\rho_0 \, (1+z)^\alpha,
\end{equation}
and the disk scale height is given as\begin{equation}
h=h_0 \, R \, (1+z)^\gamma,\label{h}
\end{equation}
where $h_0$ describes the ratio of disk height to radius at $z=0$. We note in particular that the cosmic star formation rate is known to increase towards higher redshift, thus indicating that quantities like the average gas densities in the galaxy or its scale height may be redshift-dependent as well. From these relations, the gas mass in the galaxy then follows as\begin{equation}
M\sim\rho \, R^2 \, h = \rho_0 \, h_0 \, R^3 \, (1+z)^{\alpha+\gamma}.
\end{equation}
This yields the radius as a function of gas mass and redshift,\begin{equation}
R\sim\frac{M^{1/3}}{\rho_0^{1/3} \, h_0^{1/3} \, (1+z)^{(\alpha+\gamma)/3}}.
\end{equation}
For the gas surface density, we therefore find that\begin{eqnarray}
\Sigma\sim h \, \rho&=&h_0 \, R \, (1+z)^\gamma \, \rho_0 \, (1+z)^\alpha\nonumber\\
&=&\rho_0 \, h_0 \, (1+z)^{\alpha+\gamma} \, \frac{M^{1/3}}{\rho_0^{1/3} \, h_0^{1/3} \, (1+z)^{(\alpha+\gamma)/3}}\nonumber\\
&=&\rho_0^{2/3} \, h_0^{2/3} \, (1+z)^{2(\alpha+\gamma)/3} \, M^{1/3}.\label{Sigma}
\end{eqnarray}
We assume that star formation in these galaxies is regulated by a typical Kennicutt-Schmidt-law, relating the star formation surface density $\Sigma_{\mathrm{SFR}}$ to the gas surface density $\Sigma$ as\begin{equation}
\Sigma_{\mathrm{SFR}}=\tilde{C} \, \Sigma^N \, (1+z)^{\tilde{\delta}},\label{Kenni}
\end{equation}
with $N\sim1-2$ { and $\tilde{C}$ a factor providing the normalization}. The factor $(1+z)^{\tilde{\delta}}$ allows for a potential dependence of this relation on redshift, for instance due to an evolution of the average metallicity in such galaxies. The injection rate of turbulent energy by supernovae per unit surface, which we expect to provide the dominant contribution, is then given as
\begin{equation}
\Sigma_{\mathrm{inj}}=C \, \Sigma^N \, (1+z)^\delta,\qquad C=\tilde{C} \, f_{\mathrm{mas}} \, \epsilon \, E_{\mathrm{SN}}.
\end{equation}
{ The factor $C$ provides here the normalization for the injection rate of turbulent supernova energy. We note that our results below depend only on the ratio of $C/\tilde{C}$. In this derivation,} we assumed instantaneous injection of the turbulent energy, corresponding to an average over timescales of $\sim100$~Myrs. We note that $f_{\mathrm{mas}}\sim\frac{8\%}{M_\odot}$ denotes the mass fraction of stars yielding core-collapse supernova explosions, $\epsilon\sim5\%$ denotes the fraction of supernova energy deposited as turbulence, and $E_{\mathrm{SN}}\sim10^{51}$~erg the typical energy per supernova explosion. We further note that we adopted a power-law index $\delta\neq\tilde{\delta}$ in the redshift-dependent factor, as the redshift-dependence of such turbulent energy injection is not necessarily identical to the redshift dependence of star formation. We note that, if the initial mass function (IMF) of the stars becomes more top-heavy at high redshift, as suggested for instance by studies of \citet{Omukai05} and \citet{Clark09}, some of these parameters may change, and a larger amount of far-infrared radiation per supernova explosion may occur \citep{MurphyIMF}. The IMF thus provides an additional factor which may alter this correlation. We will however consider a constant IMF here for simplicity.

In order to estimate the turbulent velocity, we assume an approximate equilibrium between turbulent energy injection and turbulent decay. With the injection rate of turbulent supernova energy per unit surface, $\Sigma_{\mathrm{inj}}$, and the turbulence decay rate per unit surface, given as the surface density of the turbulent energy divided by the characteristic Eddy timescale $h/v_t$, we have
\begin{equation}
\frac{\Sigma v_t^2}{h/v_t}=\Sigma_{\mathrm{inj}}=C \, \Sigma^N \, (1+z)^\delta.
\end{equation}
The turbulent energy is thus given as\begin{equation}
v_t=C^{1/3} \, h^{1/3} \, \Sigma^{(N-1)/3} \, (1+z)^{\delta/3}.
\end{equation}
Inserting expressions (\ref{h}) and (\ref{Sigma}) yields\begin{eqnarray}
v_t&=&C^{1/3} \, h_0^{2N/9} \, \rho_0^{(2N-3)/9} \, M^{N/9} \nonumber\\
&\times& (1+z)^{(2\alpha N+2\gamma N-3\alpha+3\delta)/9}\label{vt}
\end{eqnarray}
Due to the efficient magnetic field amplification via the small-scale dynamo, saturation can be expected within a few to a few dozen eddy-turnover times \citep{Kazantsev68, Brandenburg05, Schleicher10c, Schober12b, Beresnyak12, Schleicher13}. We note here that the presence of the dynamo in supernova-driven turbulence was already demonstrated by \citet{Balsara04} and \citet{Balsara05} with numerical simulations. For practical purposes and on the timescales considered here (i.e. $\sim100$~Myrs), we can thus assume that saturation already occured, and the magnetic energy corresponds to a fraction $f_{\mathrm{sat}}$ of the turbulent energy. We  have\begin{equation}
\frac{B^2}{8\pi}=f_{\mathrm{sat}} \, \frac{\Sigma}{h} \, v_t^2,
\end{equation}
with $\Sigma/h\sim\rho$. Using (\ref{Sigma}), (\ref{h}) and (\ref{vt}) yields\begin{eqnarray}
B&\sim& f_{\mathrm{sat}}^{0.5} \, \sqrt{8\pi} \, \rho_0^{1/2} \, (1+z)^{\alpha/2} \, C^{1/3} \, h_0^{2N/9} \, \rho_0^{(2N-3)/9} \nonumber\\
&\times&M^{N/9}(1+z)^{(2\alpha N+2\gamma N-3\alpha+3\delta)/9}.\label{BM}
\end{eqnarray}
Solving (\ref{Sigma}) for $M^{1/3}$ and inserting\begin{equation}
\Sigma=\left( \frac{\Sigma_{\mathrm{SFR}}}{\tilde{C} \, (1+z)^{\tilde{\delta}}} \right)^{1/N}
\end{equation}
from the Kennicutt-Schmidt-relation (\ref{Kenni}), we obtain\begin{eqnarray}
M^{1/3}&=&\Sigma_{\mathrm{SFR}}^{1/N} \, \tilde{C}^{-1/N} \, \rho_0^{-2/3} \, h_0^{-2/3} \nonumber\\
&\times&(1+z)^{-2(\alpha+\gamma)/3-\tilde{\delta}/N}.\label{Mthird}
\end{eqnarray}
Inserting (\ref{Mthird}) into (\ref{BM}) yields\begin{equation}
B\sim\sqrt{f_{\mathrm{sat}}8\pi} \, \rho_0^{1/6} \, \left( \frac{C}{\tilde{C}} \right)^{1/3} \, \Sigma_{\mathrm{SFR}}^{1/3} \, (1+z)^{\alpha/6+(\delta-\tilde{\delta})/3}.\label{BSFR}
\end{equation}
At $z=0$, our calculation thus yields $B\propto \Sigma_{\mathrm{SFR}}^{1/3}$. We note that the latter is compatible with the observed relation \begin{equation}
B\propto\Sigma_{\mathrm{SFR}}^{0.34\pm0.08}
\end{equation}
for spiral galaxies \citep{NiklasBeck97} and the observed relation \begin{equation}
B\propto\Sigma_{\mathrm{SFR}}^{0.25\pm0.06}
\end{equation}
for dwarf irregulars \citep{Chyzy11}. It is straightforward to show that also the normalization is consistent. Adopting a typical ISM density $\rho_0\sim10^{-24}$~g~cm$^{-3}$, a saturation level of $f_{\mathrm{sat}}\sim5\%$, as expected for supersonic turbulence \citep{FederrathPRL} and a star formation rate of $\Sigma_{\mathrm{SFR}}\sim0.1$~M$_\odot$~kpc$^{-2}$~yr$^{-1}$, we obtain a field strength of $\sim12$~$\mu$G.
Inserting the Kennicutt-Schmidt-relation (\ref{Kenni}) in (\ref{BSFR}), one can further show that\begin{equation}
B\sim\sqrt{f_{\mathrm{sat}}8\pi} \, \rho_0^{1/6} \, C^{1/3} \, \Sigma^{N/3} \, (1+z)^{\alpha/6+\delta/3}.\label{BSigma}
\end{equation}
With typical values of $N\sim1-2$, we thus find that $B\propto\Sigma^{0.33-0.66}$. This is again compatible with the observed relation in spiral galaxies, given as \citep{NiklasBeck97}\begin{equation}
B\propto\Sigma^{0.48\pm0.05}
\end{equation}
as well as in dwarf irregulars, given as \citep{Chyzy11}\begin{equation}
B\propto\Sigma^{0.47\pm0.09}.
\end{equation}
We further note that both relations (\ref{BSFR}) and (\ref{BSigma}) do not depend on the parameters $h_0$ and $\gamma$, describing the ratio $h/R$ and its evolution with redshift. A variation of these parameters from galaxy to galaxy is thus not going to influence our results. The parameter $\rho_0$, on the other hand, appears in our result, but only to the power $1/6$. When the surface density $\Sigma$ is already fixed, the additional dependence on the typical ISM density is thus rather weak, and is expected to introduce only a scatter or minor deviations from the power-law behavior derived here. The latter should not be confused with the characteristic evolution of the magnetic field strength $B\propto \rho^{1/2}$ inferred by \citet{Crutcher99}, where no constraint was imposed on the surface density.

From (\ref{BSFR}), we  expect only a minor redshift dependence concerning the relation of magnetic field strength to star formation. As the power-law indices $\tilde{\delta}$ and $\delta$ are likely similar, $(\delta-\tilde{\delta})/3$ is expected to be small, yielding a redshift evolution that is difficult to infer. A potentially stronger impact may result from the evolution of the typical ISM density with redshift, with a parameter $\alpha\sim3$ in the most optimistic case. In that case, we expect a redshift dependence of $(1+z)^{\alpha/6}$. Such a dependence could be potentially inferred if the far-infrared -- radio correlation is observed for a significant sample of galaxies at $z\sim4-6$ with the Square Kilometre Array. While a direct detection nevertheless remains difficult, we will show below that it may nevertheless influence the redshift range where the far-infrared -- radio correlation is valid.

As a result of the calculation performed here, we thus expect a correlation between star formation rate and magnetic field strength also at high redshift. Such a correlation may give rise to a (modified) version of the far-infrared - radio correlation at high redshift. In case of equipartition between magnetic and cosmic ray energy, the radio emission scales as $B^{3+\alpha_{\mathrm{syn}}}$, with $\alpha_{\mathrm{syn}}$ the synchrotron spectral index (defined as $S_{\mathrm{syn},\nu} \propto \nu^{-\alpha_{\mathrm{syn}}}$, where $S_{\mathrm{syn},\nu}$ is the flux density at frequency $\nu$), while the far-infrared luminosity per unit surface should scale as the star formation surface density $\Sigma_{\mathrm{SFR}}$. One may thus expect a relation for the ratio of surface brightness as\begin{eqnarray}
\frac{f_{\mathrm{Radio}}}{f_{\mathrm{FIR}}}&\propto& B^{3+\alpha_{\mathrm{syn}}} \Sigma_{\mathrm{SFR}}^{-1}\nonumber\\
&\propto& \Sigma_{\mathrm{SFR}}^{\alpha_{\mathrm{syn}}/3}(1+z)^{(3+\alpha_{\mathrm{syn}})(\alpha/6+(\delta-\tilde{\delta})/3)}.\label{ratio}
\end{eqnarray}
With a typical synchrotron spectral index $\alpha_{\mathrm{syn}}\sim0.9$, this ratio would scale as $\Sigma_{\mathrm{SFR}}^{0.3}$, implying a dependence on the star formation rate, similar to that of the magnetic field (\ref{BSFR}), and hence a non-linear far-infrared -- radio correlation, as claimed already by \citet{NiklasBeck97}. From Eq.~(\ref{ratio}), we see in particular that for increasing ISM densities at high redshift ($\alpha>0$), we expect an increasing magnetic field strength, and thus increasing radio emission. This is consistent with the results reported by \citet{Murphy09} as well as \citet{Ivison10a, Ivison10c} for the evolution of $q_{IR}$.

The radio -- infrared ratio (\ref{ratio}) may further be altered by the competition of different energy loss mechanisms for the cosmic ray electrons, but is roughly consistent with the observed ratios of surface brightness in far-infrared and radio. We further see that this ratio may evolve with redshift, while a correlation is in fact maintained. The latter assumes, however, that no other energy loss mechanism for cosmic ray electrons starts to become dominant. The validity of this assumption is discussed below. Based on a comparison of the different energy loss timescales, we therefore assess when such a correlation is expected to break down.

\begin{figure}[htbp]
\begin{center}
\includegraphics[scale=0.5]{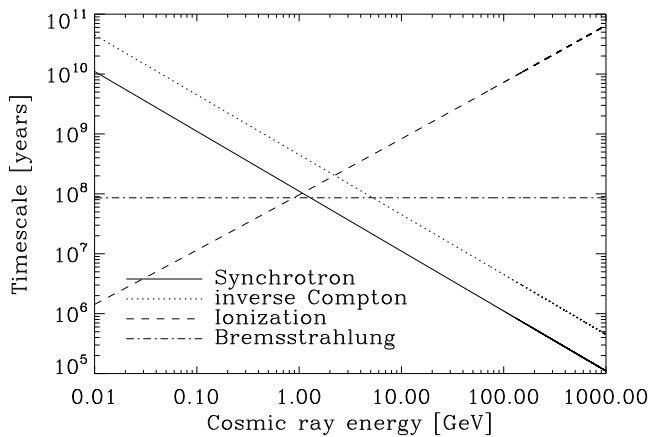}
\caption{The timescale for different cooling losses in years (synchrotron, { inverse} Compton, ionization and bremsstrahlung) vs the cosmic ray energies in GeV at $z=0$ for a magnetic field strength of $10\ \mu$G, an ISM density of $1$~cm$^{-3}$ and a radiation field of $10^{-12}$~erg~cm$^{-3}$ as typical for the Milky Way. The timescales are evaluated at the frequency $\nu_c$ corresponding to the peak synchrotron emission. We find here that synchrotron losses dominate above energies of $\sim1$~GeV.  }
\label{timescale_vs_E}
\end{center}
\end{figure}

\begin{figure}[htbp]
\begin{center}
\includegraphics[scale=0.5]{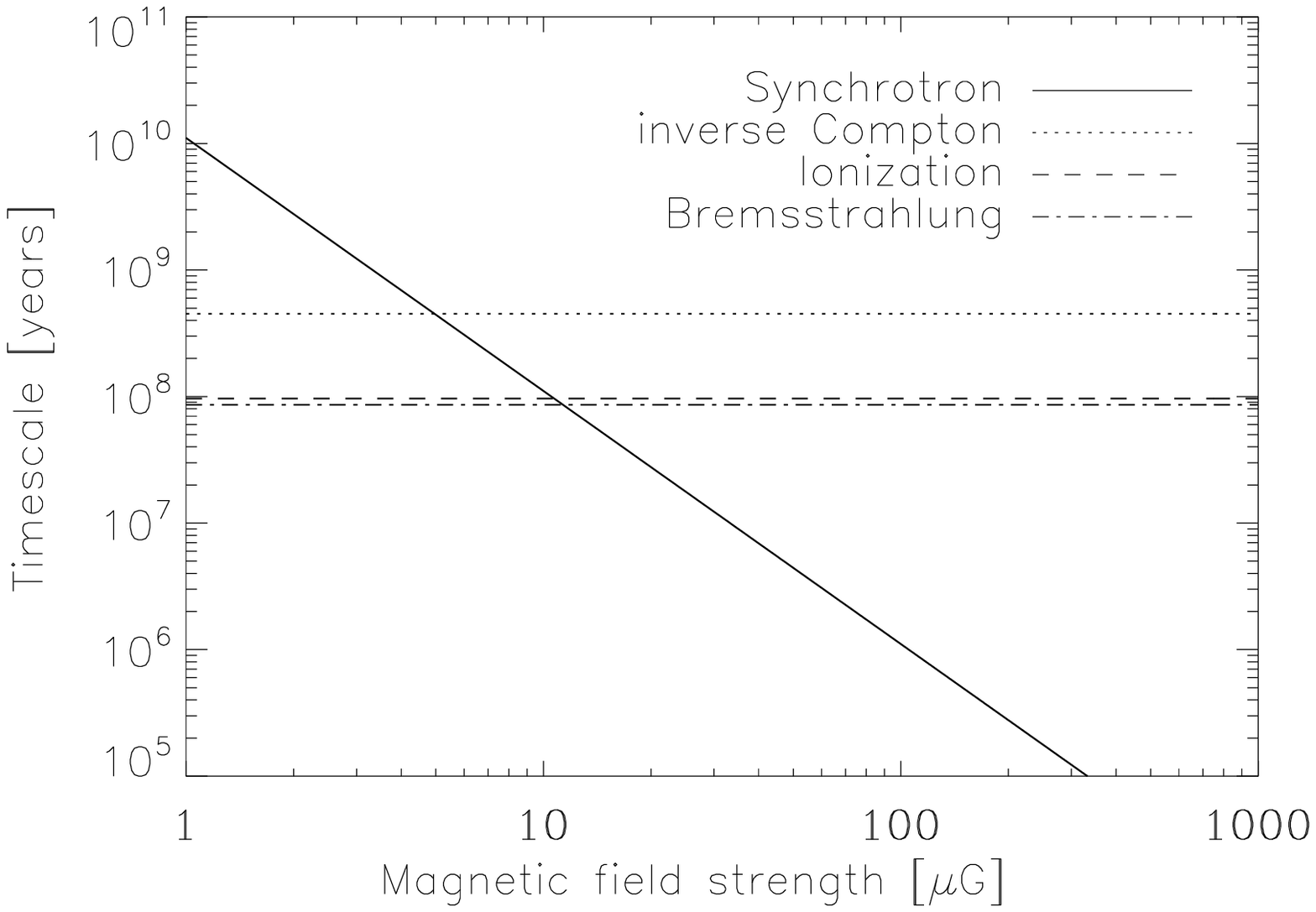}
\caption{The timescale for different cooling losses in years (synchrotron, { inverse} Compton, ionization and bremsstrahlung) as a function of magnetic field strength in $\mu$G at $z=0$ for a cosmic ray energy of $1$~GeV, an ISM density of $1$~cm$^{-3}$ and a radiation field of $10^{-12}$~erg~cm$^{-3}$ as typical for the Milky Way. The timescales are evaluated at the frequency $\nu_c$ corresponding to the peak synchrotron emission. We find here that synchrotron losses dominate above magnetic field strengths of $\sim10$~$\mu$G.  }
\label{timescale_vs_B}
\end{center}
\end{figure}

\begin{figure}[htbp]
\begin{center}
\includegraphics[scale=0.5]{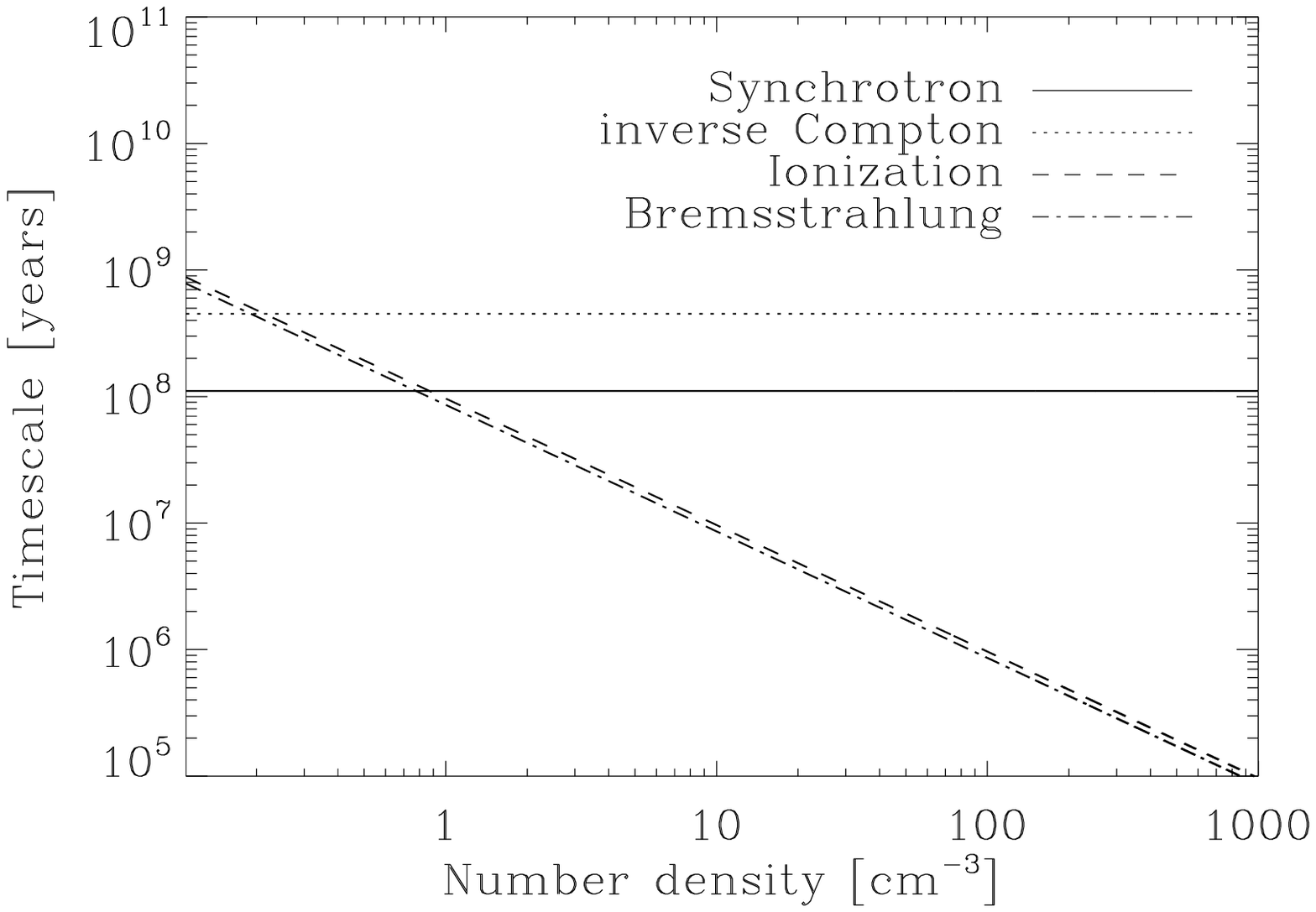}
\caption{The timescale for different cooling losses in years (synchrotron, invere Compton, ionization and bremsstrahlung) as a function of the ISM density in cm$^{-3}$ at $z=0$ for a cosmic ray energy of $1$~GeV, a magnetic field strength of $10$~$\mu$G and a radiation field of $10^{-12}$~erg~cm$^{-3}$ as typical for the Milky Way. The timescales are evaluated at the frequency $\nu_c$ corresponding to the peak synchrotron emission.  }
\label{timescale_vs_n}
\end{center}
\end{figure}

\section{The far-infrared -- radio correlation at high redshift}
We will subsequently assume that the derived relation between magnetic field strength and star formation rate holds at least approximately for high-redshift galaxies. Given this assumption, we ask whether also the far-infrared -- radio correlation (or a modified version of this relation) is expected to hold at high redshift. In particular, if cosmic ray electrons predominantly loose their energy  not by synchrotron emission but other processes, the radio emission could be strongly suppressed. In the following, we follow \citet{Longair94} and \citet{Murphy09} in the description of the physical processes, and assess under which conditions the correlation can be expected to hold.

\subsection{Energy losses of cosmic ray electrons}
As the observed non-thermal radio emission from galaxies is a result of synchrotron emission from cosmic rays in magnetic fields, we need to assess whether such synchrotron emission still remains the dominant energy loss mechanism at high redshift. In case other processes start to dominate, a considerably smaller fraction of the cosmic ray energy is available to drive synchrotron emission, implying a significant change or likely a breakdown in the far-infrared -- radio correlation. 

In the following, we start with a description of the main physical processes. Cosmic ray electrons with energy $E$ radiate in a magnetic field of strength $B$ at an effective frequency
\begin{equation}
\left(\frac{\nu_{\mathrm{obs}}}{\mathrm{GHz}}\right)=\frac{1.6\cdot10^{-2}}{(1+z)}\left( \frac{B}{\mu G} \right)\left( \frac{E}{\mathrm{GeV}} \right)^2,\label{nuobs}
\end{equation}
where $\nu_{\mathrm{obs}}$ denotes the frequency at redshift $0$, for a photon emitted at redshift $z$ with $\nu_c=\nu_{\mathrm{obs}}(1+z)$.
The cooling timescale via synchrotron emission is then given as\begin{equation}
\frac{\tau_{\mathrm{syn}}}{\mathrm{yr}}=1.4\cdot10^9\left(\frac{\nu_c}{\mathrm{GHz}}\right)^{-1/2}\left( \frac{B}{\mu G} \right)^{-3/2}.\label{syn}
\end{equation}
Energy losses may further occur via inverse Compton scattering with the radiation field in the galaxy or the photons in the cosmic microwave background. Their timescale is given as\begin{eqnarray}
\left( \frac{\tau_{ic}}{\mathrm{yr}} \right)&=&5.7\cdot10^7\left(\frac{\nu_c}{\mathrm{GHz}}  \right)^{-1/2}\left( \frac{B}{\mu G} \right)^{1/2}\nonumber\\
&\times&\left( \frac{U_{\mathrm{rad}}+U_{\mathrm{CMB}}}{10^{-12}\ \mathrm{erg}~\mathrm{cm}^{-3}} \right)^{-1},\label{iC}
\end{eqnarray}
where $U_{\mathrm{rad}}$ denotes the radiation field due to the stars in the galaxy and $U_{\mathrm{CMB}}$ the radiation field due to the cosmic microwave background. Normalizing the stellar radiation field in terms of the Milky Way, we will in the following employ \begin{equation}
U_{\mathrm{rad}}\sim10^{-12} \frac{\mathrm{erg}}{\mathrm{cm}^3} \frac{\Sigma_{\mathrm{SFR}}}{(0.3M_\odot\ \mathrm{kpc}^{-2}\ \mathrm{yr}^{-1})}\label{Urad}
\end{equation}
 and $U_{\mathrm{CMB}}\sim4.2\cdot10^{-13}(1+z)^4$~erg~cm$^{-3}$. Additional energy losses may occur via ionization on a timescale of\begin{eqnarray}
\left(\frac{\tau_{ion}}{\mathrm{yr}}\right)&=&4.1\cdot10^9\left(\frac{E}{\mathrm{GeV}}\right)\left(\frac{n_{\mathrm{ISM}}}{\mathrm{cm}^{-3}} \right)^{-1}\nonumber\\
&\times&\left( 3\ln\left( \frac{E}{\mathrm{GeV}} \right)+42.5 \right)^{-1},\label{ion}
\end{eqnarray}
for a given ISM number density $n_{\mathrm{ISM}}$, and via bremsstrahlung on a timescale \begin{equation}
\left(\frac{\tau_{\mathrm{brems}}}{\mathrm{yr}}\right)=8.6\cdot10^7\left( \frac{n_{\mathrm{ISM}}}{\mathrm{cm}^{-3}} \right)^{-1}.\label{brems}
\end{equation}
We are aware that additional cooling losses may occur from adiabatic cooling via galactic winds, even though we assume that such processes will affect only a fraction of the galaxy, and may only change the amount of synchrotron emission by a factor of a few. We further assume here that the cosmic rays are confined to the galaxy due to their magnetic fields, i.e. the timescale for escape is always larger than those of the loss processes discussed above. A violation of this condition could trigger a breakdown of the far-infrared -- radio correlation already at low redshift, which however appears not to be observed \citep{Murphy09}. Detailed modeling indicates that escape losses can be significant for galaxies with very low $\Sigma$ and also for extreme starburst galaxies with very fast galactic winds \citep{Lacki10}.

A comparison of the timescales discussed above is given in Fig.~\ref{timescale_vs_E}, where they are plotted as a function of the cosmic ray electron energy, assuming a magnetic field strength of $10\ \mu$G, an ISM density of $1$~cm$^{-3}$ and a radiation field of $10^{-12}$~erg~cm$^{-3}$ as typical for the Milky Way. In Fig.~\ref{timescale_vs_B}, we show these timescales as a function of the magnetic field strength, employing a cosmic ray energy of $1$~GeV, finding that synchrotron emission dominates above field strengths of $10$~$\mu$G. The impact of varying ISM densities is explored in Fig.~\ref{timescale_vs_n}, finding that synchrotron emission dominates at $n_{ISM}<1$~cm$^{-3}$ for the parameters explored here. It can be shown, however, that synchrotron losses again become dominant when more energetic cosmic rays are considered.

While there is  some uncertainty in the precise values of these timescales, it is important to note that synchrotron emission dominates in a realistic range of parameters, as the far-infrared -- radio correlation is established in the local Universe. In the following, we will now aim to explore the validity of this relation also at high redshift.

\begin{figure}[htbp]
\begin{center}
\includegraphics[scale=0.5]{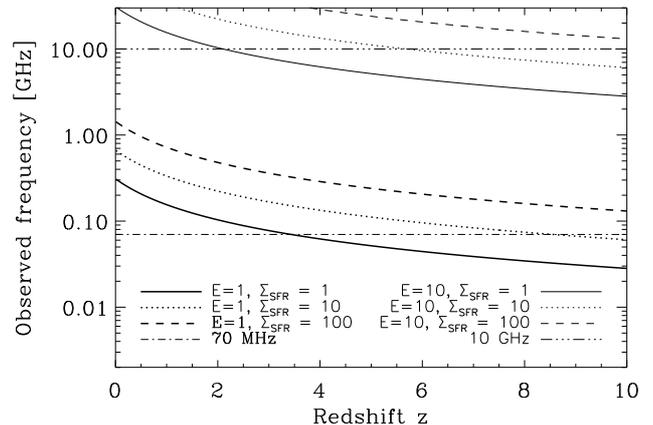}
\caption{The effective frequency of the expected synchrotron emission in the observers frame as a function of redshift for different star formation rates and cosmic ray energies. For comparison, we indicate the highest ($10$~GHz) and lowest frequencies ($70$~MHz) accessible via phase 1 and 2 of the SKA. { The cosmic ray energies $E$ are in units of GeV, while $\Sigma_{\rm SFR}$ is given in units of M$_\odot$~yr$^{-1}$~kpc$^{-2}$.} We assume here no evolution of the $B-\Sigma_{\mathrm{SFR}}$-relation with redshift, i.e. $\xi=\alpha/6=0$.  }
\label{fig:obsfreq}
\end{center}
\end{figure}

\begin{figure}[htbp]
\begin{center}
\includegraphics[scale=0.5]{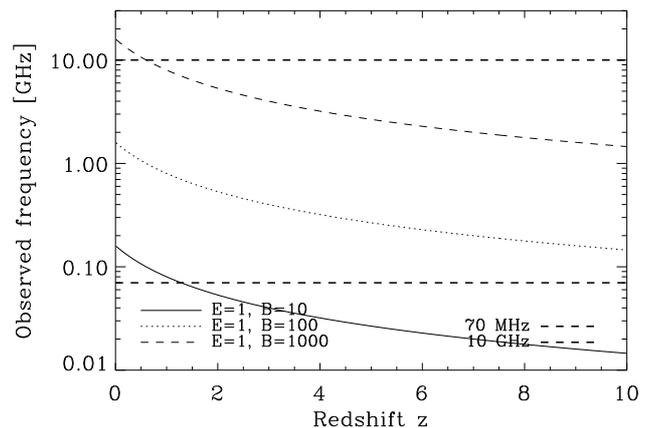}
\caption{The effective frequency of the expected synchrotron emission in the observers frame as a function of redshift for different magnetic field strength $B$ (in $\mu$G) at fixed cosmic ray energies of $1$~GeV. The highest ($10$~GHz) and lowest frequencies ($70$~MHz) accessible via phase 1 and 2 of the SKA are indicated via dashed lines.  }
\label{fig:obsfreqb}
\end{center}
\end{figure}

\begin{figure}[htbp]
\begin{center}
\includegraphics[scale=0.5]{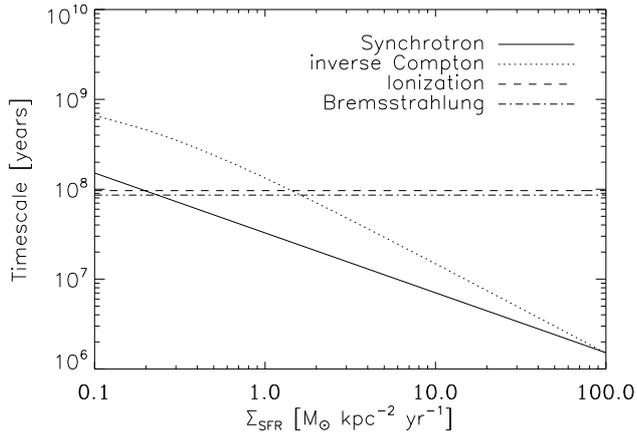}
\caption{The timescale of the different energy loss mechanisms (in years) as a function of star formation surface density at $z=0$, assuming a cosmic ray energy of $1$~GeV.  }
\label{fig:timescale_vs_sfr}
\end{center}
\end{figure}

\begin{figure}[htbp]
\begin{center}
\includegraphics[scale=0.5]{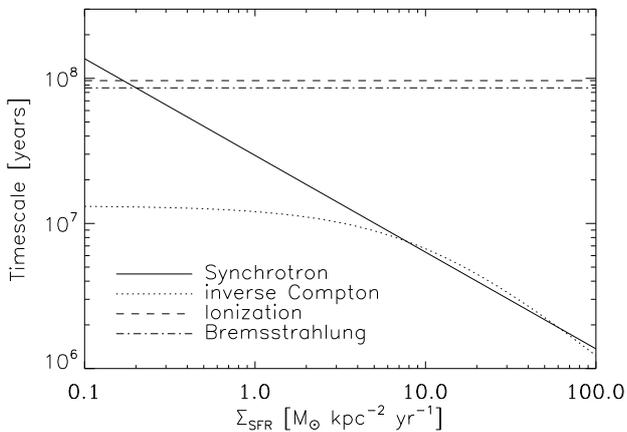}
\caption{The timescale of the different energy loss mechanisms (in years) as a function of star formation surface density at $z=2$, assuming a cosmic ray energy of $1$~GeV. We assume here no evolution of the $B-\Sigma_{\mathrm{SFR}}$-relation with redshift, i.e. $\xi=\alpha/6=0$. }
\label{fig:timescale_vs_sfr_z2}
\end{center}
\end{figure}

\begin{figure}[htbp]
\begin{center}
\includegraphics[scale=0.5]{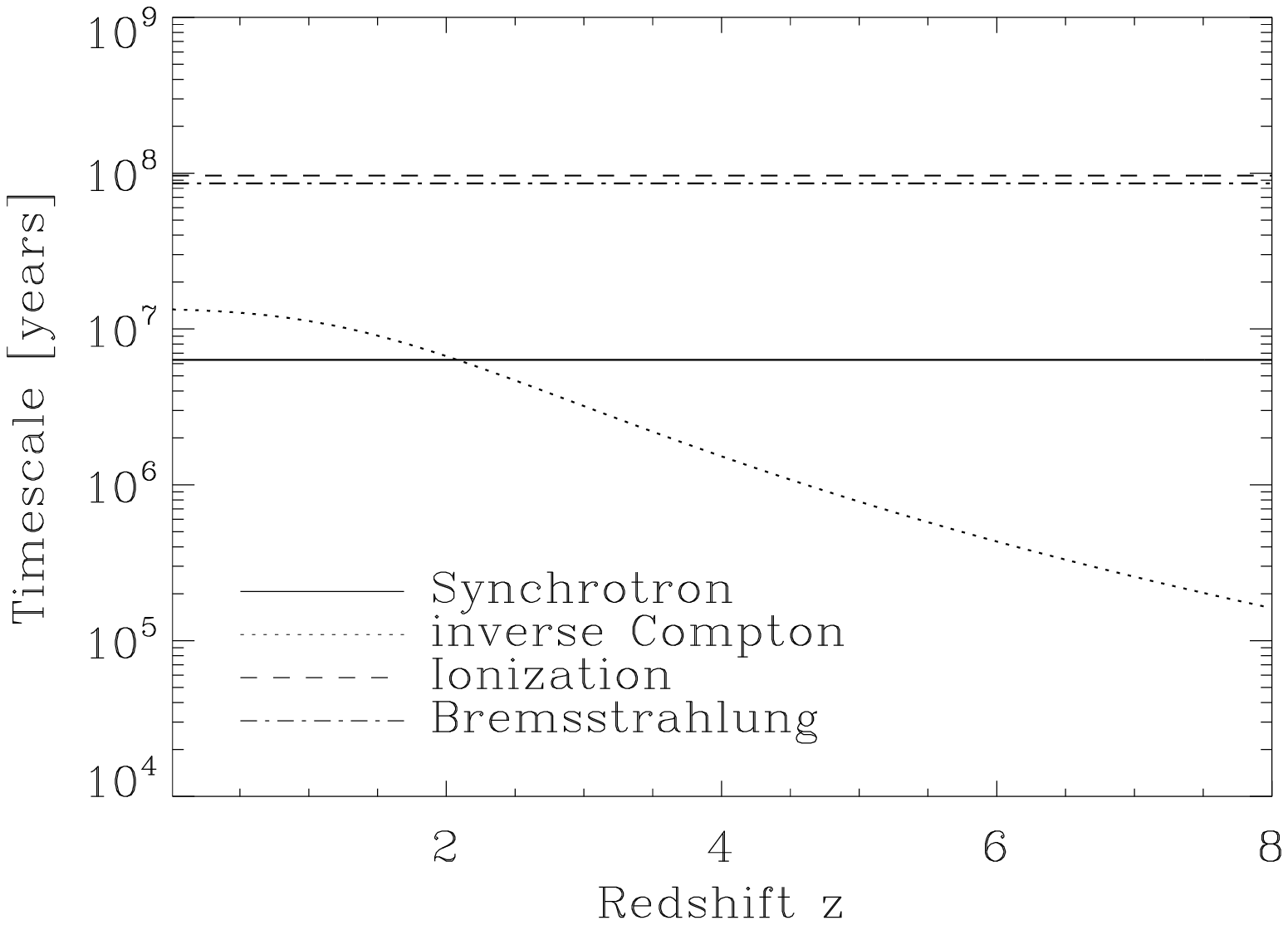}
\caption{The timescale of the different energy loss mechanisms (in years) as a function of redshift, adopting a star formation surface density of $10$~M$_\odot$~kpc$^{-2}$~yr$^{-1}$ and a cosmic ray energy of $1$~GeV. Inverse Compton losses become dominant from $z\sim2$, while synchrotron losses dominate at lower redshift. We assume here no evolution of the $B-\Sigma_{\mathrm{SFR}}$-relation with redshift, i.e. $\xi=\alpha/6=0$. }
\label{fig:e1sfr10}
\end{center}
\end{figure}

\begin{figure}[htbp]
\begin{center}
\includegraphics[scale=0.5]{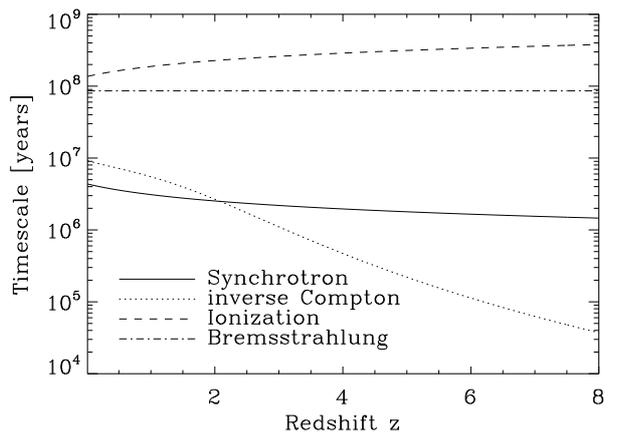}
\caption{The timescale of the different energy loss mechanisms (in years) as a function of redshift, adopting  a star formation surface density of $10$~M$_\odot$~kpc$^{-2}$~yr$^{-1}$ and an observed frequency of $1.4$~GHz. Inverse Compton losses become dominant from $z\sim2$, while synchrotron losses dominate at lower redshift. We assume here no evolution of the $B-\Sigma_{\mathrm{SFR}}$-relation with redshift, i.e. $\xi=\alpha/6=0$. }
\label{fig:constfreq}
\end{center}
\end{figure}

\subsection{Implications for synchrotron emission at high redshift}

After the description of the main physics in the previous subsection, we now assess the potential implications for synchrotron emission at high redshift. In order to probe the far-infrared -- radio correlation at high redshift, a central condition is that the effective frequency of the synchrotron emission is within or close to the observable frequency range. We adopt here the frequency range planned for phase 1 and 2 of the SKA, ranging from $70$~MHz to $10$~GHz.\footnote{ SKA phase 1 will be constructed from 2017-2020, with initially $10\%$ of the total collecting area. During phase 2 (envisaged construction from 2022-2025), the collecting area is then completed to $100\%$.} We further employ the observed relation between magnetic field strength and star formation rate, $B\propto\Sigma_{\mathrm{SFR}}^{1/3}$, and calculate the radiation field strength from the star formation rate following (\ref{Urad}). Initially, we will assume a constant ISM density of $n_{ISM}=1$~cm$^{-3}$, but will explore deviations from this scenario below. Adopting a generic cosmic ray energy of $1$~GeV, we can then plot the frequency of the observed emission as a function of redshift for different star formation rates and cosmic ray energies using  Eq.~(\ref{nuobs}) (see Fig.~\ref{fig:obsfreq}). Considering only the synchrotron frequency, it appears that one can always find a cosmic ray energy between $1$ and $10$~GeV such that the frequency lies within the frequency range of SKA. For comparison, we also show the effective frequency at a fixed cosmic ray energy of $1$~GeV for a set of different magnetic field strength (Fig.~\ref{fig:obsfreqb}). In this case, we inverted the relation between magnetic field strength and star formation rate, so that the radiation field in the galaxy can still be inferred via Eq.~(\ref{Urad}).

However, we also need to check whether the timescale for synchrotron emission is still the smallest timescale for the energy losses, as the radio emission can be significantly suppressed if other energy losses dominate. For this purpose, we calculate the timescale for the energy losses at $z=0$ and $z=2$ as a function of star formation rate (see Fig.~\ref{fig:timescale_vs_sfr} and \ref{fig:timescale_vs_sfr_z2}). At $z=0$, we find that synchrotron emission dominates up to star formation surface densities of $100$~M$_\odot$~kpc$^{-2}$~yr$^{-1}$, where it becomes comparable to the inverse Compton losses. The latter can be understood, as the synchrotron losses scale as $B^{-3/2}$, while inverse Compton scales as $B^{1/2}$, and $B\propto\Sigma_{\mathrm{SFR}}^{1/3}$.

At $z=2$, however, we find that the synchrotron timescales are approximately equal to the inverse Compton timescales for $\Sigma_{\mathrm{SFR}}>6$~M$_\odot$~kpc$^{-2}$~yr$^{-1}$, while the inverse Compton scattering via the CMB dominates at low star formation rates and weak magnetic fields. At higher star formation rates, the dominant radiation field is produced by the galaxy, but due to the enhanced magnetic field both timescales are now comparable. At that point, inverse Compton losses may already start to alter the far-infrared -- radio correlation  at low star formation rates. The same is indicated in Fig.~\ref{fig:e1sfr10}, where the energy loss timescales are plotted as a function of redshift for a star formation surface density of $10$~M$_\odot$~kpc$^{-2}$~yr$^{-1}$ and a fixed cosmic ray energy of $1$~GeV. For comparison, we show the evolution of the timescales  at a fixed observed frequency in Fig.~\ref{fig:constfreq}. In both cases, the inverse Compton timescale becomes shorter than the timescale for synchrotron emission at $z\sim2$.

\subsection{The breakdown of the far-infrared -- radio correlation at high redshift}

As indicated above, inverse Compton losses start to dominate over synchrotron losses from $z\sim2$, thus yielding a potential breakdown of the far-infrared -- radio correlation at high redshift. In this section, we assess in more detail how the breakdown of this relation may depend on the properties of the galaxy population, as well as the intrinsic relation between magnetic field strength and star formation rate. In order to derive the point when synchrotron losses start to dominate, we assume that inverse Compton scattering is due to the cosmic microwave background, and equate Eqs.~(\ref{syn}) and (\ref{iC}), yielding a critical magnetic field strength as a function of redshift:\begin{equation}
\left( \frac{B_c}{\mu \mathrm{G}} \right)\sim3.2 \, (z+1)^2.\label{Bc}
\end{equation}
We thus note that synchrotron emission is expected to dominate for all field strengths above this value. While in case of high star formation rates, a radiation field stronger than the CMB will be present, these are also expected to yield larger magnetic field strengths, which will likely compensate. We note that the critical magnetic field strength is independent of frequency, as both processes show the same frequency dependence.

\begin{figure}[htbp]
\begin{center}
\includegraphics[scale=0.5]{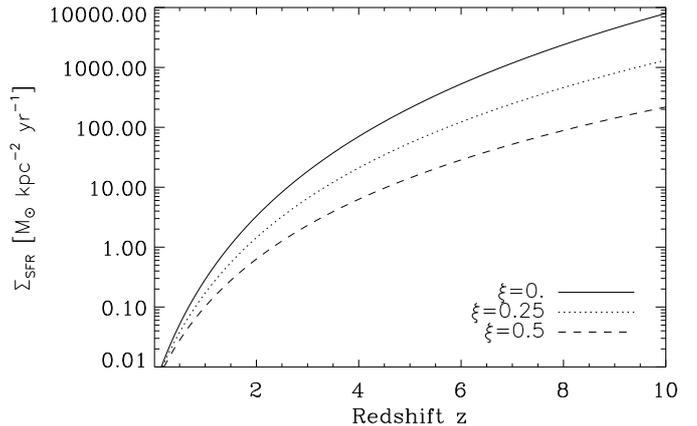}
\caption{The critical star formation rate as a function of redshift for different values of $\xi=\alpha/6$, where the parameter $\alpha$ describes the evolution of typical ISM densities with redshift, i.e. $n_{\mathrm{ISM}}=n_0(1+z)^\alpha$. We assume here that observations are pursued at the effective frequency as given in Eq.~(\ref{nuobs}). }
\label{sfr_vs_z}
\end{center}
\end{figure}

\begin{figure}[htbp]
\begin{center}
\includegraphics[scale=0.5]{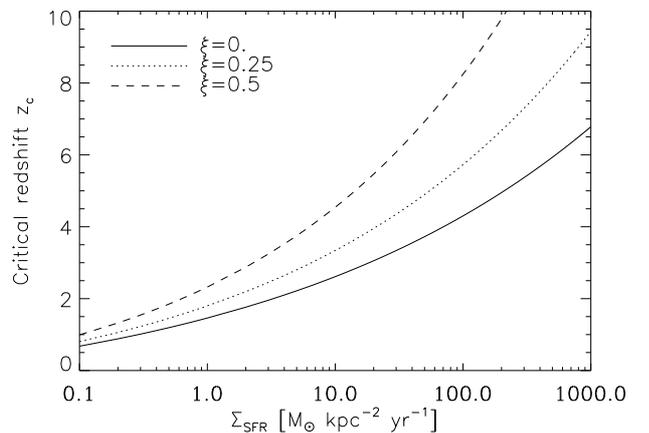}
\caption{The redshift where the far-infrared -- radio correlation breaks down as a function of star formation rate, for different values of $\xi=\alpha/6$, where the parameter $\alpha$ describes the evolution of typical ISM densities with redshift, i.e. $n_{\mathrm{ISM}}=n_0(1+z)^\alpha$. We assume here that observations are pursued at the effective frequency as given in Eq.~(\ref{nuobs}).   }
\label{zvssfr}
\end{center}
\end{figure}

In order to translate the critical field strength derived here into galaxy properties, we employ a relation \begin{equation}
B\sim9\,\mu \mathrm{G}\left(\frac{\Sigma_{\mathrm{SFR}}}{0.1\ M_\odot\ \mathrm{kpc}^{-2}\ \mathrm{yr}^{-1}}  \right)^{1/3}(1+z)^\xi
\end{equation}
as motivated in section~\ref{interpret}. We recall that the main redshift dependence may result from the redshift dependence of the mean ISM density, $n_{\mathrm{ISM}}=n_0(1+z)^\alpha$, yielding $\xi=\alpha/6$. Inserting this relation in (\ref{Bc}) thus yields a critical star formation rate, given as\begin{equation}
\Sigma_{\mathrm{SFR,c}}\sim0.0045 \, (1+z)^{6-3\xi} \, M_\odot\ \mathrm{kpc}^{-2}\ \mathrm{yr}^{-1}.\label{SFRz}
\end{equation}
For star formation surface densities above this value, galaxies are thus expected to lie on the far-infrared -- radio correlation, while radio emission will be strongly suppressed for considerably smaller star formation rates. At redshift zero, for a typical spiral galaxy with a stellar disk of $\sim10$~kpc, this critical value corresponds to a typical star formation rate of $1.4$~M$_\odot$~yr$^{-1}$. From the scaling in Eq.~(\ref{SFRz}), it however increases rapidly with redshift, yielding a critical value of $200-1000$~M$_\odot$~yr$^{-1}$ for a galaxy of the same size at $z=2$, corresponding to infrared-luminosities of 
$\sim2\times10^{12}-1\times10^{13}$~L$_\odot$. 

For comparison, we note that the sample explored by \citet{Ivison10a} contains luminous and ultraluminous galaxies with luminosities of $10^{11}-10^{13}$~L$_\odot$ at $z=2$. While it is interesting that these are comparable to the characteristic value derived here, we stress that the critical quantity is however the star formation surface density, which we cannot directly infer from the data. Such an analysis can however be pursued for spatially resolved galaxies, in particular from lensed systems, such as SMMJ~02399-0136 at $z=2.8$ \citep{Frayer98, Ivison10b}. With a total star formation rate of $\sim10^3$~M$_\odot$~yr$^{-1}$ and a spatial extend of the CO-disk of $25$~kpc, we obtain a star formation surface density of $\sim2$~M$_\odot$~kpc$^{-2}$~yr$^{-1}$, which is { comparable and higher} to the critical { star formation surface density} depending on $\xi$. With at least $10^8$~M$_\odot$ of molecular gas observed in this region, we can however savely assume that the typical ISM density here is higher than for present-day galaxies, thus implying a decreased critical threshold. From this example, we thus conclude that both the star formation rate as well as the ISM density are relevant.

The evolution of the critical star formation surface density derived in Eq.~(\ref{SFRz}) is illustrated in Fig.~\ref{sfr_vs_z} for different values of $\xi$. While it provides effectively no constraint at low redshift, critical star formation surface densities above $100$~M$_\odot$~kpc$^{-2}$~yr$^{-1}$ are required for the standard scenario ($\xi=0$) at $z\sim6$. We note that the critical values are somewhat decreased for $\xi>0$. In turn, we can thus determine the redshift where the breakdown of the correlation occurs for a given star formation surface density. Solving (\ref{SFRz}) for $z+1$, we obtain\begin{equation}
z_c+1=\left( \frac{\Sigma_{\mathrm{SFR}}}{0.0045\ M_\odot\ \mathrm{kpc}^{-2}\ \mathrm{yr}^{-1}} \right)^{1/(6-3\xi)}.\label{critz}
\end{equation}
The relation is shown in Fig.~\ref{zvssfr} for different values of $\xi$. We note that it holds at all frequencies, as the inverse Compton and synchrotron timescales have the same frequency dependence. While for $\xi=0$, the relation will break down already at $z_c\sim4$, we note that even a moderate value of $\xi=0.25$ or $\alpha=1.5$ would be sufficient to maintain the relation until $z_c\sim6$. We summarize the redshift where this break-down occurs as a function of star formation rate and $\xi$ in Table~1.

\begin{table}[htdp]
\label{critred}
\begin{center}
\begin{tabular}{c|c|c}
$\xi$ & $\Sigma_{SFR}$~[M$_\odot$~kpc$^{-2}$~yr$^{-1}$] & $z_c$\\
\hline
0 & 1 & 1.5\\
0 & 10 & 2.7\\
0 & 100 & 4.4\\
0.25 & 1 & 1.8\\
0.25 & 10 & 3.3\\
0.25 & 100 & 5.7\\
$0.5$ & 1 & 2.3\\
0.5 & 10 & 4.5\\
0.5 & 100 & 8.2
\end{tabular}
\end{center}
\caption{The redshift where the far-infrared -- radio correlation is expected to break down due to inverse Compton scattering, as a function of star formation rate and evolution parameter $\xi=\alpha/6$.}
\end{table}

In order to assess whether additional energy losses may become relevant, we further compare the synchrotron timescale to the timescale for bremsstrahlung emission. For this purpose, we equate (\ref{syn}) and (\ref{brems}), employing $n_{\mathrm{ISM}}=n_0(1+z)^\alpha$. As these processes have a different frequency dependence, we can now derive a critical frequency, below which bremsstrahlung dominates, of\begin{eqnarray}
\nu_{\mathrm{obs,brems}}&=&\frac{\nu_{\mathrm{c,brems}}}{1+z}\nonumber\\
&=&0.27\ \mathrm{GHz}\, n_0^2 \, (1+z)^{2\alpha-1}\left(\frac{B}{10\mu G} \right)^{-3}.
\end{eqnarray}
 This condition however provides no strong constraint, as the SKA will observe even at frequencies of $10$~GHz. While it is trivially fulfilled at low redshift, the behavior at high redshift depends on the  value of $\alpha$. As the critical magnetic field strength due to inverse Compton losses however scales as $(1+z)^2$, the effect will never dominate in case of high-frequency observations.  It however needs to be taken into account for observations at constant frequencies, which we discuss below, in particular for weak fields and at low frequencies (as low as $70$~MHz for the SKA).

In a similar fashion, we can equate the timescales for ionization and synchrotron losses, (\ref{syn}) and (\ref{ion}), yielding a critical frequency of\begin{eqnarray}
\nu_{\mathrm{obs,ion}}&=&\frac{\nu_{\mathrm{c,ion}}}{1+z}\nonumber\\
&=&0.29\,\mathrm{GHz} \, n_0^2 \, (1+z)^{2\alpha-1} \, \left(\frac{E}{1\,\mathrm{GeV}}\right)^{-2}\nonumber\\
&\times&\left(\frac{B}{10\,\mu\mathrm{G}}\right)^{-3}.
\end{eqnarray}
It is again trivial to see that the condition yields no additional constraint for a reasonable parameter space, as the critical magnetic field strength due to inverse Compton losses already scales as $(1+z)^2$. As we discuss below, the condition may only become relevant for weak fields at low redshift and low frequencies.

\begin{figure}[htbp]
\begin{center}
\includegraphics[scale=0.5]{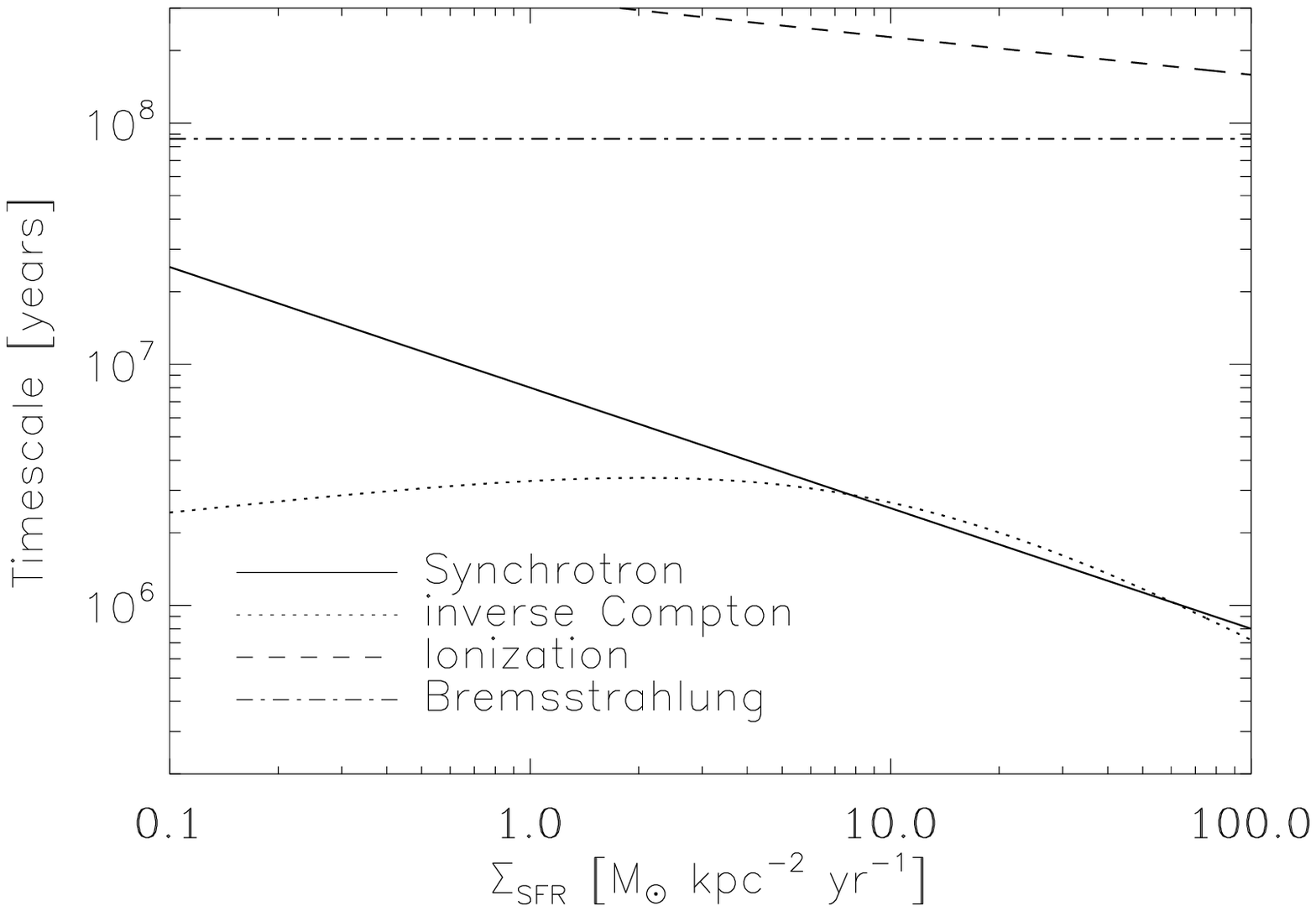}
\caption{The timescale of the different energy loss mechanisms (in years) as a function of star formation surface density at $z=2$, assuming a cosmic ray energy of $1$~GeV. We assume here that the observations are performed at a fixed frequency of $1.4$~GHz, and no evolution of the characteristic ISM density with redshift ($\xi=\alpha/6=0$). We find that synchrotron emission will be strongly suppressed for star formation rates below $\sim8$~M$_\odot$~yr$^{-1}$~kpc$^{-2}$. }
\label{fig:timescale_vs_sfr_constnu}
\end{center}
\end{figure}

\begin{figure}[htbp]
\begin{center}
\includegraphics[scale=0.5]{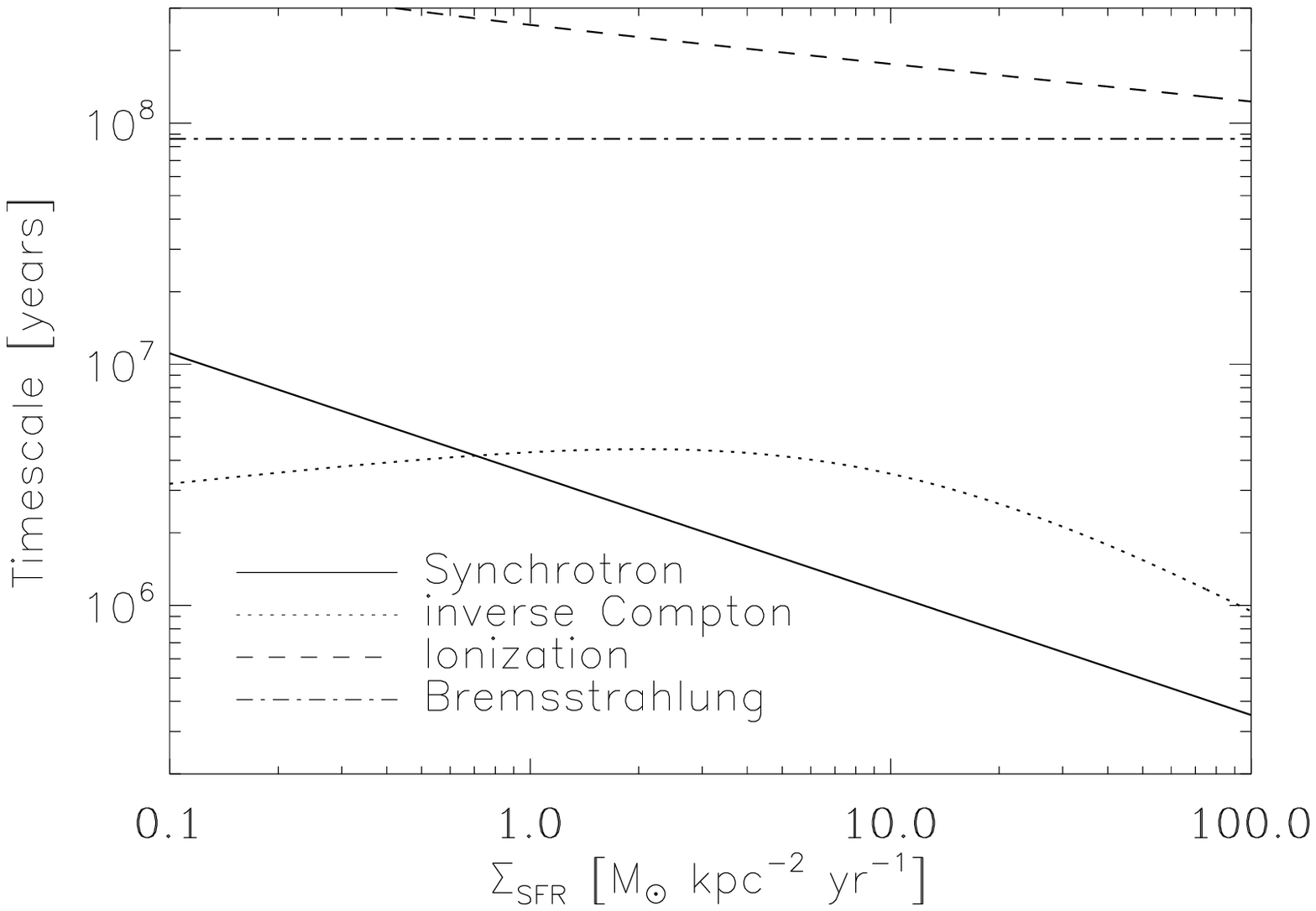}
\caption{The timescale of the different energy loss mechanisms (in years) as a function of star formation surface density at $z=2$, assuming a cosmic ray energy of $1$~GeV. We assume here that the observations are performed at a fixed frequency of $1.4$~GHz, and a strong evolution of the characteristic ISM density with redshift ($\xi=\alpha/6=1/2$). In this case, strong synchrotron emission is possible down to star formation rates of $\sim1$~M$_\odot$~yr$^{-1}$~kpc$^{-2}$.  }
\label{fig:timescale_vs_sfr_alpha}
\end{center}
\end{figure}

\subsection{Observations at constant frequency}

We finally discuss the implications of observations at constant frequency. For this purpose, we first note that the critical redshift for the breakdown of synchrotron emission, Eq.~(\ref{critz}) is independent of observed frequency, as the timescales for synchrotron emission and inverse Compton scattering have the same frequency dependence. Indeed, plotting the timescales for the characteristic energy losses at $z=2$ as a function of star formation rate for $\xi=\alpha/6=0$, we find that strong synchrotron emission is only possible for star formation rates above $\sim8$~M$_\odot$~yr$^{-1}$~kpc$^{-2}$ (see Fig.~\ref{fig:timescale_vs_sfr_constnu}), in agreement with our previous results. Similarly, we note that the emission is strongly enhanced if the relation between magnetic field strength and star formation evolves with redshift, as indicated in Fig.~\ref{fig:timescale_vs_sfr_alpha} for $\xi=\alpha/6=1/2$. These calculations are pursued for an observed frequency of $1.4$~GHz.

However, the timescale for ionization losses depends on the cosmic ray energy, and for different observed frequencies, the dominant radio emission will result from different cosmic ray energies E. Solving Eq.~(\ref{nuobs}) for E, we have\begin{equation}
\left( \frac{E}{\mathrm{GeV}} \right)=\sqrt{\left( \frac{\nu_{\mathrm{obs}}}{\mathrm{GHz}} \right)\frac{1+z}{1.6\cdot10^{-2}}\left( \frac{\mu\mathrm{G}}{B} \right)}.\label{Eobs}
\end{equation}
Requiring that the timescale for synchrotron emission is shorter than for ionization losses and assuming a fixed observed frequency, we find the following condition for synchrotron emission to be dominant:
\begin{equation}
(1+z)^{\alpha-1}<23.2\left( \frac{\nu_{\mathrm{obs}}}{\mathrm{GHz}} \right)^{1/2}  \left( \frac{B}{\mu\mathrm{G}} \right)\left( \frac{\mathrm{cm}^{-3}}{n_0} \right)      
\end{equation}
For $\alpha<1$ (little / no density evolution with redshift), this condition is almost trivially fulfilled. For $\alpha>1$, it yields a weak constraint, which is however less relevant than the constraint from inverse Compton emission.

\begin{figure}[htbp]
\begin{center}
\includegraphics[scale=0.5]{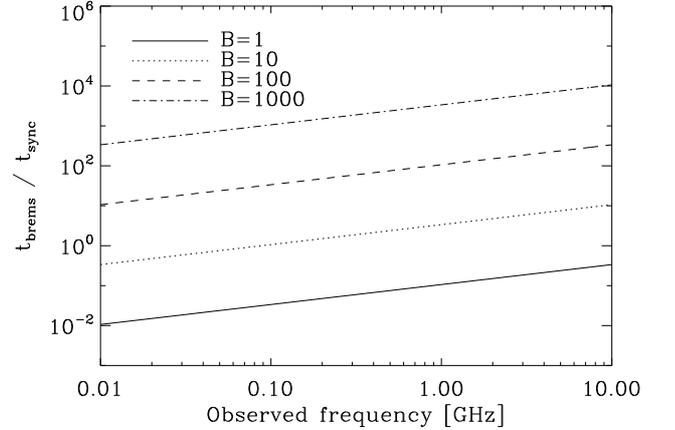}
\caption{Ratio of the timescales for bremsstrahlung losses over synchrotron losses at $z=2$, as a function of observed frequency for different magnetic field strength. The calculation assumes a generic ISM number density of $n_{\mathrm{ISM}}=1$~cm$^{-3}$. In this case, bremsstrahlung losses only become relevant for moderate fields below $\sim10$~$\mu$G. We note, however, that this ratio scales as $n_{\mathrm{ISM}}^{-1}$, implying that bremsstrahlung becomes more relevant in dense galaxies. }
\label{fig:ratio}
\end{center}
\end{figure}

In a similar fashion, one may check whether bremsstrahlung losses may dominate during the evolution. Requiring that the timescale for synchrotron emission is shorter than for bremsstrahlung emission, again adopting a constant observed frequency, we find the relation\begin{equation}
(1+z)^{\alpha-1/2}<0.061 \left( \frac{\nu_{\mathrm{obs}}}{\mathrm{GHz}} \right)^{1/2}  \left( \frac{B}{\mu\mathrm{G}} \right)^{3/2}\left( \frac{\mathrm{cm}^{-3}}{n_0} \right)  
\end{equation}
For $\alpha<1/2$, the condition is again trivially fulfilled. At $\alpha>1/2$, the condition translates into a constraint on the magnetic field strength, given as\begin{equation}
B>6.45\,\mu\mathrm{G}\left( \frac{n_0}{\mathrm{cm}^{-3}} \right)^{2/3} \left( \frac{\nu_{\mathrm{obs}}}{\mathrm{GHz}} \right)^{-1/3} (1+z)^{2\alpha/3-1/3}.
\end{equation}
The behavior is demonstrated in Fig.~\ref{fig:ratio}, where we show the ratio of the timescales for bremsstrahlung losses over synchrotron losses at $z=2$ as a function of frequency. While for a moderate ISM number density of $n_{\mathrm{ISM}}=1$~cm$^{-3}$, bremsstrahlung losses only become relevant for weak magnetic fields below $\sim10$~$\mu$G, this ratio scales as $n_{\mathrm{ISM}}^{-1}$, implying stronger constraints for dense galaxies. In these cases, observations at high frequencies above $1$~GHz are particularly favorable to probe high-redshift magnetic fields.

\section{Summary and discussion}

We have shown that the observed relation between star formation rate and magnetic field strength can be understood as a result of turbulent magnetic field amplification, where turbulence is injected by supernova explosions. This model implicitly assumes that turbulent magnetic field amplification efficiently occurs on short timescales, as previously demonstrated in numerous studies \citep[e.g.][]{Kazantsev68, Brandenburg05, Beresnyak12, Schober12b, Schleicher13, Bovino13}. Effectively, the magnetic field strength is thus always close to saturation, and reflects the amount of turbulence present in the galaxy.

Due to the efficiency of this amplification process, such a relation between star formation rate and magnetic field strength is expected even at high redshift. If the typical ISM density is increasing with redshift, we expect stronger magnetic fields and initially thus stronger radio emission. The latter is consistent with observational results by \citet{Murphy09} and \citet{Ivison10a, Ivison10c}, who report a radio excess for high-redshift systems. Such an evolution indicates a modification, but not a breakdown of the far-infrared -- radio correlation. { It is further instructive to compare our theoretical expectations to the sample of \citet{Cram98}, who explored star formation in faint radio galaxies in a sample of more than $700$ local galaxies. These authors compared the star formation rate derived from the radio flux, SFR$_{\rm radio}$, with the star formation rates derived from the U-band, SFR$_{\rm U-band}$, and derived from H$\alpha$, SFR$_{H\alpha}$. For SFR$_{\rm radio}<1$~M$_\odot$~yr$^{-1}$, the values of SFR$_{\rm U-band}$ and SFR$_{H\alpha}$ were typically enhanced by an order of magnitude, indicating that radio emission is suppressed at low star formation rates, as we propose here. At the same time, they also report that SFR$_{\rm U-band}$ and SFR$_{H\alpha}$ become smaller than SFR$_{\rm radio}$ at SFR$_{\rm radio}>1$~M$_\odot$~yr$^{-1}$. The latter implies enhanced radio fluxes for high star formation rates, and is consistent with a non-linear evolution of the far-infrared - radio correlation, as proposed by \citet{NiklasBeck97} and consistent with Eq.~(\ref{ratio}) in our model. Of course, one should note that our predictions strictly apply only to the star formation surface density rather than the star formation rate, but the correlation appears nevertheless to be consistent.} 

{ At high redshift, \citet{Pannella09} provided a stacking analysis of radio data for a large sample of galaxies at $z=1-3$, comparing the UV-derived (uncorrected) star formation rates to the star formation rate inferred from the radio data. Considering that the dust attenuation has not been accounted for, the results appear roughly consistent, with star formation rates of $10-100$~M$_\odot$~yr$^{-1}$. They also confirm the correlation between star formation rate and stellar masses, as previously reported by \citet{Rodighiero01}. As their sample still includes many galaxies around redshift $1$, the latter currently provides no strong constraint to our model. If this trend is however confirmed from fainter galaxies or higher redshift, it may confirm our hypothesis that the mean ISM density is increasing with redshift.
}

{ An additional factor that may influence the evolution of the far-infrared - radio correlation is the evolution of the mean metallicity and the IMF.} We have shown in Eq.~\ref{BSFR}) that the  magnetic field strength will scale as\begin{equation}
\left( \frac{C}{\tilde{C}} \right)^{1/3}=\left(f_{\mathrm{mas}} \epsilon E_{\mathrm{SN}} \right)^{1/3}.
\end{equation}
In case of less efficient cooling, a smaller amount of the supernova energy could be radiated away, thus increasing the  fraction of energy $\epsilon$ going into turbulence. For the energy of supernova explosions, we expect them to be constant to a reasonable degree, although more detailed calculations concerning the supernova explosions of low metallicity stars may be necessary to understand this point. Overall, an additional increase of the magnetic energy may also result from this factor, even though an improved understanding of low-metallicity star formation and the deposition of their supernova feedback will be required for a detailed assessment. As we demonstrate in our model, the evolution of the relation between star formation rate and magnetic field strength does not depend on the ratio of disk scale height to radius, but has a weak dependence on the evolution of the ISM number density with redshift, described as $n_{ISM}=n_0(1+z)^\alpha$. 

On the other hand, such a correlation between magnetic field strength and star formation surface density  does not necessarily imply that also the far-infrared -- radio correlation still holds at high redshift. In particular, we show here that  inverse Compton scattering may become the dominant energy loss mechanism for cosmic rays in high-redshift galaxies, implying that radio emission could become strongly suppressed and galaxies should become bright in X-ray emission. Specifically, we find that at each redshift, a critical magnetic field strength of at least $B_c\sim3.2\ (z+1)^2\mu$G is required for synchrotron emission to occur. For weaker fields, we expect an actual breakdown of the correlation.

In the framework of our model, the latter translates into a critical star formation rate, given as $\Sigma_{\mathrm{SFR,c}}\sim0.0045\ (1+z)^{6-3\xi} M_\odot\ \mathrm{kpc}^{-2}\ \mathrm{yr}^{-1}$. On the other hand, considering galaxies with a given star formation surface density $\Sigma_{\mathrm{SFR}}$, we have calculated the redshift $z_c=\left( \Sigma_{\mathrm{SFR}}/(0.0045\ M_\odot\ \mathrm{kpc}^{-2}\ \mathrm{yr}^{-1}) \right)^{1/(6-3\xi)}$  where the far-infrared -- radio correlation is expected to break down. 

In this model, the critical star formation surface density at redshift zero corresponds to a star formation rate of $1.4$~M$_\odot$~yr$^{-1}$ for a typical spiral galaxy of $10$~kpc radius. The latter is for instance comparable to the star formation rate in the Milky Way system. At $z=2$, the critical star formation rate is already $200-1000$~M$_\odot$~yr$^{-1}$ for a galaxy of the same size at $z=2$, or infrared-luminosities of $\sim2\times10^{12}-1\times10^{13}$~L$_\odot$. These luminosities are in fact comparable to the typical luminosities in the sample by \citet{Ivison10a, Ivison10c}. It is however plausible that high redshift galaxies will be more dense and compact, implying both an increased star formation surface density and ISM density. Only a direct measurement of these quantities, as pursued for instance for SMMJ~02399-0136 \citep{Frayer98, Ivison10b} allows us to directly test our predictions. Especially with  ALMA\footnote{ALMA webpage: http://www.almaobservatory.org}, we however expect that such resolved star formation surface densities can be provided for a larger sample. With such a sample, the model proposed here can be tested by exploring the far-infrared -- radio correlation for fixed star formation surface densities as a function of redshift. As shown here, the correlation will ultimately break down when inverse Compton scattering off CMB photons starts to dominated over synchrotron emission. Assuming a star formation surface density of $10\ M_\odot\ \mathrm{kpc}^{-2}\ \mathrm{yr}^{-1}$, this breakdown is expected at $z\sim2.7$ in our standard scenario with $\xi=0$. If, on the other hand, the ISM density increases significantly with redshift ($\xi=0.5$), this point can be delayed until $z\sim4.5$. When the break-down is measured, X-ray observations may provide an alternative means of probing magnetic field evolution due to inverse Compton scattering at even higher redshift.

At low frequencies, bremsstrahlung losses may provide an additional constraint on the observable field strength, depending on the evolution of the mean ISM densities as a function of redshift. At the same time, we expect them to provide no severe constraint at frequencies of $1-10$\,GHz, thus still allowing to probe magnetic fields out to the critical redshift described above.

{ Additional care must be taken to account for potential biases in the observational determination of  $q_{FIR}$ and $q_{IR}$ and their evolution with redshift. In particular, a selection of galaxies in the IR band and taking into account only sources with corresponding radio detections may bias the sample to the radio-bright galaxies. In this respect, \citet{DelMoro13} suggested that no evolution in these parameters occurs if contributions from hidden AGN are taken into account \citep[see also][]{Daddi07}. Limitations due to potential biases can however be overcome with an accurate stacking analysis, and by accounting for sources where only the upper limits on the radio flux have been obtained \citep{Gruppioni03}. These uncertainties have also been acknowledged by \citet{Ivison10a, Ivison10c}, and further investigations are thus required to disentangle the contributions to the radio flux  and accurately infer the evolution of the far-infrared - radio correlation with redshift.
}

Of course,  our model is still subject to simplifications, and more sophisticated scenarios may need to be considered. For instance, additional turbulence can be injected by accretion or mergers, potentially leading to stronger magnetic fields and hence deviations or scatter around the relation proposed here. Simulations following the cosmic evolution of galaxies with their magnetic fields would thus be valuable to assess how the relation between star formation and magnetic fields evolves over time. At the same time, microphysical processes as the cosmic ray diffusion need to be more reliably assessed at high redshift, in order to explore whether such additional losses may become relevant. In this way, probing the far-infrared -- radio correlation provides a pathway of probing the physics of high-redshift galaxies, including star formation, magnetic fields and cosmic rays.

\begin{acknowledgements}
We thank Eric Murphy and Olaf Wucknitz for a careful reading and valuable feedback on our manuscript.  DRGS thanks for funding from the {\em Deutsche Forschungsgemeinschaft} (DFG) in the {\em Schwerpunktprogramm} SPP 1573 ``Physics of the Interstellar Medium'' under grant SCHL 1964/1-1, and via the SFB 963/1 ``Astrophysical flow instabilities and turbulence'' (project A12). RB thanks for funding from DFG FOR1254 ``Magnetisation of interstellar and intergalactic media''. { We further thank the anonymous referee for valuable comments that helped to improve our manuscript.}
\end{acknowledgements}


\end{document}